\documentclass[aip,reprint]{revtex4-1}
\usepackage{graphics}
\usepackage{graphicx}
\usepackage[english]{babel}
\usepackage{epstopdf}

\draft % marks overfull lines with a black rule on the right

\begin{document}

%\preprint{}

%\title{In-situ liquid crystal film target inserter for repeatable moderate repetition rate laser matter experiment} %Old title of paper

\title{In-situ, variable thickness, liquid crystal film target inserter for moderate repetition rate intense laser applications}



\author{P. L. Poole$^*$}
\author{C. Willis}
\author{G. E. Cochran}
\author{R. J. Hanna}
\author{C. D. Andereck}
\author{D. W. Schumacher}
\email[]{$^*$poole.134@osu.edu}

%\thanks{}
\affiliation{$^1$The Ohio State University, 191 West Woodruff Ave, \\ Columbus, OH 43210, USA \\}


\date{\today}

\begin{abstract}
Liquid crystal films have recently been demonstrated as variable thickness, planar targets for ultra-intense laser matter experiments and applications such as ion acceleration. By controlling the parameters of film formation, including liquid crystal temperature and volume, their thickness can be varied on-demand from 10 $nm$ to above 10 $\mu m$. This thickness range enables for the first time real-time selection and optimization of various ion acceleration mechanisms. Previous work employed these targets in single shot configuration, requiring chamber cycling after the pre-made films were expended. Presented here is a film formation device capable of drawing films from a bulk liquid crystal source volume to any thickness in the aforementioned range. Design parameters have been selected to optimize the device for prolonged, moderate repetition rate operation. The device will form films under vacuum within 2 $\mu m$ of the same location each time, well within the Rayleigh range of even tight $F/ \#$ systems, thus removing the necessity for realignment between shots. Additionally, film formation for several hundred shots is possible before the chamber needs to be opened. The repetition rate of the device exceeds 0.1 $Hz$ for sub-100 $nm$ films, enabling moderate repetition rate plasma target insertion for state of the art lasers currently in use or under development. Characterization tests of the device performed at the Scarlet laser facility at The Ohio State University will be presented.
\end{abstract}


%Laser-plasma interactions, 52.38 -r, Liquid crystal phase transitions in, 64.70.M-, Laser-driven acceleration 41.75.Jv

\maketitle %\maketitle must follow title, authors, abstract and \pacs

\section{Introduction}

As ultra-intense laser technology improves, new facilities are able to achieve progressively higher repetition rates. The BELLA facility at LBNL is a currently operating petawatt facility with a 1 $Hz$ repetition rate,\cite{Leemans14} and the ELI facilities under construction will have petawatt lasers at the 10 $Hz$ rate.\cite{Banerjee14} Increased shot rates promise higher flux for experiments and applications such as ion acceleration, neutron radiography, and energetic beam generation,\cite{Murnane89, Roth13, Kneip09} but only if the problems of target insertion, diagnostic operation, and data handling at higher repetition rates can be solved. Presented here is an instrument for inserting low-Z targets at sustained moderate repetition rates in the form of freely suspended films made of a commonly available liquid crystal. The basic approach described is potentially scalable to much higher repetition rates.

A successful target insertion mechanism for ultra-intense laser experiments and applications must meet multiple requirements simultaneously. First, targets are typically thin, either to permit laser interaction throughout the entire volume or to reduce the mass that laser generated radiation (e.g. electrons, ions, electromagnetic radiation) must pass through. For example, there is presently much interest in laser based ion acceleration, both for the physical mechanisms involved and for its possible applications, including ion cancer therapy.\cite{Masood14} To clearly access the full range of known acceleration mechanisms, planar target thicknesses from 10 $nm$ to above 1 $\mu m$ are required. Currently, such targets are ordered and manufactured in advance in fixed thicknesses. This is suitable for single shot experiments but not for higher repetition rate operation. Second, ultrahigh intensities in the range of 10$^{18}$ $W/cm^2$ to 10$^{22}$ $W/cm^2$ require tight focus geometries and, hence, Rayleigh ranges on the order of 10 $\mu m$ or less. Accordingly, target alignment to focus must be repeatable to within a few $\mu m$ to prevent unacceptable intensity variations from shot to shot. Third, targets must have sufficiently large extent to minimize plasma damage to the supporting structures. The required extent will depend on repetition rate and is not yet well understood, but $\sim$1 $mm$ sizes are a reasonable expected minimum. Finally, the target expense must be low to enable practical sustained runs at high repetition rates. There may be many other requirements, for example involving debris management, but the above constitute a minimal set.

Given the necessity and importance of improving high power laser target technology, a number of methods have been proposed and implemented to provide targets satisfying these requirements, or some subset, for existing and upcoming laser facilities that operate beyond the single shot level. Among them are gas and liquid sprays, ubiquitous in low intensity laser research, that allow for kHz repetition rate shots at $mJ$ pulse energies.\cite{Karsch03, Ter-Avetisyan03, Ter-Avetisyan06, Schnurer07, Henig09} These targets are non-ideal for a number of high intensity laser applications, including ion acceleration, for a few reasons: first, their dispersed nature creates a small laser-target interaction region;\cite{Karsch03, Ter-Avetisyan06} second, in the case of liquids, it is difficult to control droplet thickness or make droplets smaller than 10 $\mu m$;\cite{Ter-Avetisyan03} and lastly, also in the case of liquids, the spherical target geometry accelerates ions in three dimensions as it expands, ultimately resulting in lower ion energy and yields than for planar targets.\cite{Schnurer07, Henig09} 

A related approach using liquid jets has been demonstrated using cryogenic hydrogen\cite{Zastrau14} and water\cite{Feister14} and has found considerable success for use with x-ray lasers or low energy $kHz$ optical lasers. This class of targets has thus far been restricted to thicknesses of order 10 $\mu m$ in extent and larger, and in the case of water the relatively high vapor pressure is incompatible with ultrahigh intensity laser experiments. Additionally the target geometry is cylindrical, resulting in reduced ion energy and yield as with spherical droplet targets.

An approach that retains planar geometry is the use of tape targets.\cite{Nishiuchi08, Sokollik10} Here a thin spool of target material is passed between two motors while the laser fires in between. While these have achieved reasonable repetition rates approaching 1 $Hz$, the target thickness is fixed and thus far must be at least several $\mu m$ thick to prevent tape breakage from the laser interaction. Furthermore, the action of the motors typically results in positional jitter along the incoming laser direction on the order of tens of $\mu m$. While improvement of these technologies continues, they currently are not feasible for high repetition rate intense laser experiments.

To summarize, a number of difficult requirements must be met simultaneously for the insertion of thin targets in laser systems that fire faster than the formerly typical once per hour. To the authors' knowledge, no previously demonstrated method provides quick or easily controlled thickness variation down to sub-$\mu m$ levels. In particular, the two orders of magnitude thickness range that is paramount for ion acceleration is not readily accessible. This is a key requirement both to optimize the desired interaction to laser parameters such as energy, pulse width, and contrast, as well as to access acceleration mechanisms that rely on target thickness such as Target Normal Sheath Acceleration,\cite{Snavely00, Hatchett00} Radiation Pressure Acceleration, \cite{Esirkepov03} and Break-out Afterburner.\cite{Yin06} Target thickness scanning is a commonly used method to investigate the physics of these newly discovered acceleration mechanisms, and limitations in target creation, insertion, and alignment hinder progress in this area.

Liquid crystals, however, can be formed into membranes that preserve planar geometry while adding the considerable benefit of on-demand thickness variation between a few $nm$ and several $\mu m$.\cite{Poole14} Their low vapor pressure allows target formation and thickness manipulation at typical target chamber pressure levels, and the low volume per film renders them ideal targets for long-term high repetition rate use. 

To this end an instrument has been developed that forms liquid crystal films under vacuum with the temperature and volume control necessary to achieve the $nm$ to $\mu m$ thickness range demonstrated in recent single-shot experiments. The device consists of a vertically sliding wiper that draws a liquid crystal film as it moves across a 4 $mm$ aperture within a copper frame; as such it is called the Linear Slide Target Inserter (LSTI). The design passively maintains excellent alignment even in tight focus geometries, and has been demonstrated for fast thin film target insertion on recent experiments. Section II includes details of the design for forming and controlling the thickness of liquid crystal films, while Section III shows data taken with the device.

\section{LSTI design}

\subsection{Liquid crystal properties}

Liquid crystals exhibit states of matter incorporating features of conventional solid and liquid phases. For a given liquid crystal, these mesophases exist over certain known temperature ranges and are characterized by different levels of molecular ordering. For example, in the smectic phase molecules have positional order in that they group into roughly single-molecule-thick layers, and orientational order in that, on average, all molecules point along one direction. The liquid crystal of choice for these experiments, 4'-octyl-4-biphenyl (8CB), manifests its smectic phase between 294.5 $K$ and 306.5 $K$, and has a smectic layer thickness of 2.7 $nm$. It is possible to form smectic phase films with anywhere from several to many hundreds of layers.

8CB can be formed into freely-suspended liquid crystal films while in its smectic phase due to the surface tension inherent to the material. A film will form within a rigid frame if a small volume of liquid crystal is dispensed next to an aperture and then drawn across with a stiff wiper. To both manipulate the thickness and obtain a uniform film, temperature control over the range of a few degrees with regulation to better than 0.1 $^{\circ}C$ and volume adjustment down to 100 $nL$ are required. The details of film formation parameters to obtain thickness control between 50 $nm$ and 5000 $nm$ are described elsewhere.\cite{Poole14} Previous experiments were performed in single-shot mode by first making films in individual copper frames to the desired thickness via mechanical wiping. These films were then transferred to the vacuum chamber, which was subsequently pumped down to the 10$^{-6}$ Torr level before final target alignment. Throughout this process the film thickness did not change. This procedure mimicked solid target insertion and alignment as currently performed using metal foils, however high repetition rate shots necessitate in-vacuum film formation, and the properties of liquid crystal target production are well enough understood to enable this. To this end a film formation device was designed that could be installed on a target positioner within the experimental vacuum chamber.

 \begin{figure}
 \includegraphics[width=0.5\textwidth]{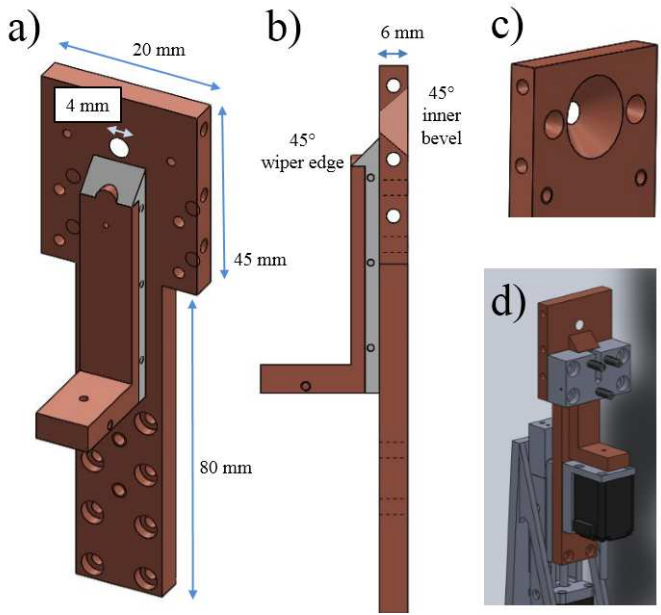}
 \caption{\label{fig:LSTIdim} a) Isometric view of Linear Slide Target Inserter frame and wiper, including the 4 $mm$ aperture for film formation, clearance holes for temperature monitoring and control through the side, and mounting holes at the bottom. b) Side view depicting the wiper and film aperture bevel angle of 45$^{\circ}$. c) Rear view of film aperture area showing the bevel that provides repeatable film formation position, as described below. d) Full assembly of LSTI including bridge piece for wiper guidance and driving motor.}
 \end{figure}

\subsection{In-situ film formation}

Schematic images of the LSTI film formation device are shown in Fig.\,\ref{fig:LSTIdim}. Since accurate temperature control is of paramount importance in producing a desired thickness, both the frame and wiper of the device are primarily made of copper. The frame has a 4 $mm$ diameter clearance hole to act as an aperture for film formation. The back side of this aperture has a 45$^{\circ}$ bevel to allow attenuated, unamplified short pulse light to be collected and characterized on an in-situ camera just behind chamber center, as well as to accommodate widely diverging beams leaving the target interaction area. This aperture bevel has the added benefit of moving a forming film to its edge, providing a repeatable film position, as will be described below. The smaller holes to the left and right of the film formation aperture are for mounting shadowgraphy alignment targets for use with the Scarlet confocal positioning system for target alignment.\cite{Willis15}

The wiper is another copper piece with a beveled top edge that can be pushed vertically with a full travel of 50 $mm$ by a NEMA 8 vacuum motor (Micronix). The wiper is held flush to the frame by a polyether ether ketone (PEEK) bridge, which also serves to guide the wiper motion. This piece has three Delrin-tipped spring-loaded plungers to provide a variable amount of force pressing the wiper down onto the frame, as firm contact is necessary for film formation.

The volume of liquid crystal present in a single film of sub-$\mu m$ thickness is on the order of 10 $nL$, so dispensing a precise volume for just one film is challenging. A 1 $mm$ diameter clearance hole is placed in the wiper for the application of liquid crystal directly to the space between the wiper and frame. This can be done with tubing connected from this clearance hole to a precision syringe pump (Harvard Apparatus) for fine control of volume deposition. In practice, it is often easier to simply apply a volume on the order of 1 $\mu L$, significantly more than that required for one film, and to vary other film formation parameters to control thickness. In this way one application of liquid crystal volume before chamber evacuation can provide hundreds of films before more is needed, in this case bypassing the need for the syringe pump and small inner diameter tubing. This enables moderate repetition rate target formation for hours at a time without requiring vacuum chamber cycling.

The initial volume deposition also acts as a lubricant that prevents scratching of the wiper and frame surfaces. With insufficient lubrication these scratches will grow over several draws into channels that change the liquid crystal volume present near the film aperture, hindering full thickness control. To minimize these surface effects both the wiper and frame pieces are polished with successively fine grains of sandpaper from 160 to 2000 grit, followed by the use of polishing compound until the copper has a mirror finish. Maintaining this level of surface smoothness increases the thickness repeatability significantly.

An additional measure taken to reduce frame scratching was to modify the wiper to include a Teflon piece affixed to the bottom of the wiper, shown in Fig.\,\ref{fig:LSTIdim}a. With this wiper design scratches only form on the Teflon wiper, and only over a large number of film draws. While the Teflon cover ensures greater film thickness control through scratch prevention for many more draws than the pure Cu wiper, the reduced adhesion of liquid crystal to the Teflon surface results in slightly different film formation behavior, as will be discussed in Section III.

Thickness control\cite{Poole14} requires precise temperature modulation and monitoring, typically around 28.0 $^{\circ}C$. For film formation at atmospheric pressure this is achieved by inserting two resistive cartridge heaters operated at low power (25 $W$ maximum) into 3 $mm$ diameter clearance holes bored at different heights horizontally through the LSTI frame. In the vacuum environment the motor acts as a heat source, resulting in a temperature gradient at the film area that produces uneven films and in general impedes thickness control. To accommodate this, the resistive heaters are removed and their clearance holes are fitted with a copper tubing line connected through the chamber wall to a small water chilling unit. A third channel allows a type K thermocouple to be inserted internal to the frame near the film formation aperture. In this way the frame temperature can be maintained to within 0.1 $^{\circ} C$ of the desired temperature regardless of the variable thermal load from the motor.

\begin{figure}
 \includegraphics[width=0.4\textwidth]{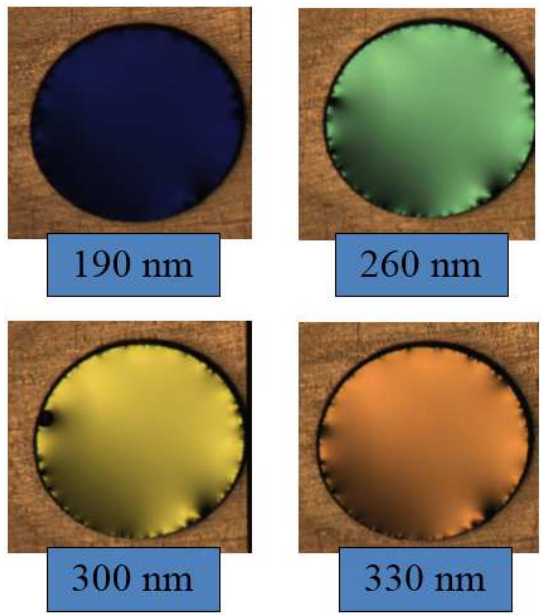}
 \caption{\label{fig:filmspread} The LSTI device has fine thickness control on the formation of liquid crystal targets. Shown are four films separated in thickness enough to show different colors, though finer thickness manipulation down to 10 $nm$ is possible with this film formation device.}
 \end{figure}

Temperature and volume control as described here ensures that a meniscus region does not form where the film attaches to the frame, resulting in a 4 $mm$ diameter film of uniform thickness (i.e. consistently the same number of smectic liquid crystal layers). Example images of LSTI films are shown in Fig.\,\ref{fig:filmspread}, where the color comes from constructive thin film interference of reflected light. The small features visible on the edge of these films arise from edge defects in the aperture, and are the only factor preventing liquid crystal films from being optically flat reflectors. Subsequent version of the LSTI will improve upon this edge design for applications such as plasma mirrors, to be discussed in a subsequent publication.

\section{Film control}

\subsection{Position repeatability}

Film thickness is monitored via a spectral reflectance measurement from a Filmetrics F-20 device: white light of a known spectrum (halogen lamp) is reflected from the target film back towards the device; modulations in the reflected spectrum from interference in the thin film along with some knowledge of the liquid crystal index of refraction are used to iteratively solve for the thickness. This process has a 50 $ms$ integration time so it can be done at high repetition rates, and has an accuracy of 2 $nm$ with proper calibration. The white light image is relayed from outside the vacuum chamber through a viewport using two achromatic lenses to provide minimal beam aberration, which ensures best thickness measurement accuracy. This light is set up to hit the target at normal incidence so that the reflected spectrum returns to the device for analysis, and has an automatic shutter to protect the input fiber during a laser shot.

Of note is the repeatability of the film formation plane of the LSTI. To test this, a digital linear micrometer is used in conjunction with Scarlet's sub-Rayleigh range confocal positioning target alignment device\cite{Willis15} to determine the film location upon destruction and subsequent reformation. A film was placed at the best focus of the confocal positioner, then a new film was drawn and the target frame was moved until it was measured to again be at best focus of the alignment device. The digital micrometer recorded the net displacement in the target plane during this $Z$ realignment. The RMS value of film position was within 2 $\mu m$, well within the Rayleigh range of the $F/2$ Scarlet laser. This test is repeated throughout the duration of experimental runs, with consistent film formation results.

\begin{figure}
 \includegraphics[width=0.5\textwidth]{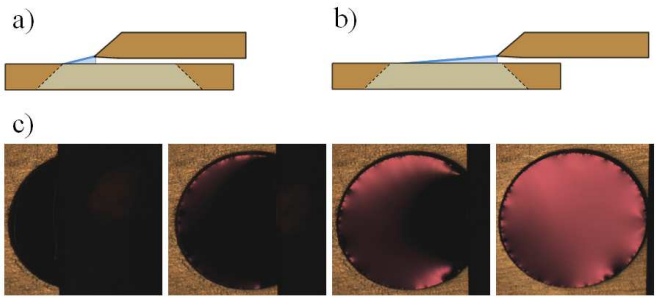}
 \caption{\label{fig:wiperangle} The geometry of the beveled aperture and raised wiper provide a consistent film formation plane. a) Cartoon depicting a film initially forming between the aperture and the wiper, such that its center area is pulled at an angle with respect to the aperture plane. b) As the wiper continues drawing, more of the film is in plane with the copper aperture, having been pulled to the front of the bevel feature. c) Snapshots during film formation. As the wiper moves left to right more area of the forming film is brought nearer to parallel with the frame aperture, resulting in the correct angle for observing the pink color stemming from constructive interference at this film thickness of 530 $nm$.}
 \end{figure}

The reason for this precision in film formation position lies in the beveled aperture machined in the copper frame, as illustrated in Fig.\,\ref{fig:wiperangle}. As the wiper is drawn down, a film initially forms with its upper edge connected to the copper frame and its lower edge still connected to the wiper tip, which is shaped during the polishing step to lie a small distance (sub-$mm$) above the plane of the copper frame. Because of this, the film initially forms at an angle with respect to the copper frame (and its eventual resting plane). Note that the forming film is always touching either the circular aperture or the wiper tip on all sides, and this angle manifests as a curvature to the overall film, as can be seen in Fig.\,\ref{fig:wiperangle}c. The surface tension of the liquid crystal is such that the film is pulled forward towards the wiper as it moves downward, pulling the liquid crystal film so it is on the extreme front edge of the aperture bevel as it is being formed. As the wiper proceeds through the down-stroke, this angle between the film center area and the copper frame plane decreases, and as the wiper edge leaves the aperture area the film snaps into place, now fully contained within the copper frame aperture. In this way films are brought to the same plane within 2 $\mu m$ of the previous film location each time. To further test this, films many $\mu m$ thick were formed using the slowest motor speeds. The location of the front surface of each film was measured to be a distance from the previous film front side by a value equal to the difference in their measured thicknesses, consistently, indicating that the back surface of each film forms in a stationary spot within the frame aperture.

This consistent film formation removes the necessity for new target alignment between shots, a problem that is currently the chief time constraint for solid target experimentation on high repetition rate systems. This feature also improves experimental down time on moderate repetition rate systems like the once per minute Scarlet laser, which has been able to collect data from solid targets for the first time at its native repetition rate because of the consistent target formation of the LSTI.

\subsection{Thickness control}

 \begin{figure}
 \includegraphics[width=0.5\textwidth]{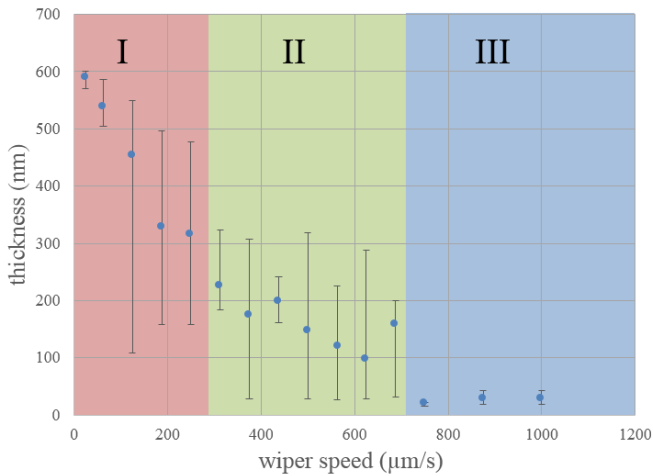}
 \caption{\label{fig:wiperscan} Control of film thickness via LSTI wiper speed. Here the vertical bars show the range of films formed over 5 draws at each speed, while the dots indicate the average of those films. The average thickness increase as the wiper speed decreases. The shaded areas indicate three different regions of thickness range, discussed in the text.}
 \end{figure}

Although temperature, applied volume, and surface polish all affect film formation, for practical long-term film repeatability wiper speed is the quickest and most convenient variable for tuning film thickness. For example, while the volume applied demonstrably governs the upper limit to thickness produced at any speed, sub-$100 nm$ films can still be produced when the applied volume is greater than 10 $\mu L$. Furthermore, while in general a higher temperature results in thicker films, even at temperatures of 29.5 $^{\circ}C$ (near the smectic/nematic phase transition) a high wiper speed will still generate a thin film. 

The effect of wiper speed can be seen in Fig.\,\ref{fig:wiperscan}, where three observed regions of film formation have been highlighted. Region III, at the highest wiper speeds, produces a sub-$100 nm$ film with precision within $10 nm$ each time. Film thickness in this region is only a weak function of the applied volume and temperature. Thus, films may be formed at high repetition rates at these wiper speeds for applications where a consistent, thin target is desired, as in the case of currently studied ion acceleration mechanisms. The surface tension inherent to the smectic liquid crystal phase (on the order of 50 $dynes/cm$ \cite{OswaldBook}) allows film formation even at quite high wiper speeds. As a result the maximum draw time is limited currently by the choice of motor, not the liquid crystal itself. The motor used for this device allows thin films to be formed at a repetition rate of about 0.3 $Hz$, but the principle of freely suspended liquid crystal film formation is scalable to higher repetition rates.

Regions I and II both have less precision at a given wiper speed, due chiefly to the difficulty of controlling the volume drawn into a film at the necessary 100 $nL$ level. The vertical bars in Fig.\,\ref{fig:wiperscan} indicate the range of film thicknesses produced at the given wiper speed, while the dot shows the average over these thicknesses. Region II occurs in the several hundred $\mu m/s$ wiper speed range, varying slightly with applied volume and temperature. Here film thicknesses are typically between 100 and 500 $nm$, changing as a slightly stronger function of volume than in Region III. The average film thickness within Region II increases as the wiper speed decreases, although with a much larger possible thickness range than in Region III. Region I is typically below 300 $\mu m/s$ wiper speed and results in film thicknesses up to several $\mu m$, where this maximum possible thickness is correlated directly to the volume of liquid crystal applied. 

The difference in wiper material manifests itself chiefly in Region III behavior: here the copper wiper produces films around 30 $nm$, while the Teflon wiper makes thicknesses of 80 $nm$. This shift in thickness produced with wiper material is probably also present in Regions I and II, but not noticeable due to the wide range of thicknesses possible at those wiper speeds. This consistency at high speeds is thought to be caused by the intrinsic flow properties of liquid crystal: at slower speeds there is more time for volume to leave the liquid crystal meniscus and form a film of greater thickness. At sufficiently high speeds there is a favorable thickness based on how the meniscus attaches to the material of the wiper. In this way a Teflon wiper produces thicker Region III films due to the increased area of attachment between meniscus and wiper.

The lack of precise thickness control is not generally a problem when thickness scans are desired since the thickness of each film is readily determined as described earlier. If a specific thickness is required for each shot, this can be achieved in one of two ways at the cost of increased time between shots. One approach is to apply exactly the correct volume from a precision syringe pump through the hole provided for this purpose, visible in Fig.\,\ref{fig:LSTIdim}. A second approach, and the one recommended for its mechanical simplicity, is to start with a thicker film and reshape it with additional draws of the wiper.

Beginning a draw from 1.5 $mm$ above the aperture will result in a thicker film than if the initial draw is from 0.5 $mm$ above the aperture, simply because of the extra volume brought into the frame aperture region. In this way the thicker range at a given wiper speed in Fig.\,\ref{fig:wiperscan} can be preferentially accessed by beginning a draw from higher up on the frame. Then the wiper can be moved again over the aperture area at a slightly higher speed, effectively wiping away some number of smectic layers from the film to reduce its thickness.

Though films formed in Regions I and II do not have the reproducible precision of those from Region III, the thickening and thinning techniques outlined above allow the desired film thickness to be approached over only a handful of wipes. By choosing the appropriate wiper speed, the desired thickness can be reached within 10 percent in only a few draws. Since each draw takes a matter of seconds, multiple re-draws are easily feasible for most current high power laser systems. For example, the LSTI and its associated thin film formation techniques enable the 400 TW Scarlet laser to achieve shots at its full repetition rate of 1/min for the first time. Additionally, the stability of the smectic liquid crystal phase combined with its low vapor pressure enables films formed within the LSTI to maintain their thickness nearly indefinitely, making them ideal as well for laser systems with repetition rates of once every few minutes to hours.

\section{Conclusion}

The in-situ thin film target positioner presented here offers liquid crystal film formation in vacuum with on-demand thickness variation from 10 $nm$ to over 10 $\mu m$ at a repetition rate as high as 0.3 $Hz$ for very thin films and better than one shot per minute in general. The design allows for consistent film formation location, removing the necessity for alignment between shots. The device uses the liquid crystal 8CB, and so is ideal for high intensity laser experiments that rely on low-$Z$ targets, such as ion acceleration. The low vapor pressure, low volume per film, and wide thickness range make targets formed with this device ideal for many high intensity, moderate repetition rate laser applications. Films formed with this device are of uniform thickness across the aperture due to careful control of temperature, liquid crystal volume, and wiper speed as detailed here. Additionally, the use of wiper speed as a control mechanism enables large volumes of liquid crystal to be used, such that several hundred films can be formed with this device during chamber pumpdown, further increasing data collection rates over conventional solid target insertion techniques. Extension of the mechanical wiping technique described here is possible to much higher repetition rates, and will be pursued in future work.

\section*{Acknowledgment}

We would like to thank Xiao-lun Wu of the University of Pittsburgh, Hiroshi Yokoyama and Peter Palffy-Muhoray of the Kent State University Liquid Crystal Institute, and Cheol Park and Kyle Meienberg from the University of Colorado for fruitful discussions, as well as R.R. Freeman of The Ohio State University and Michael Storm for their contributions to this project. This work was supported by the DARPA PULSE program through a grant from AMRDEC and by the NNSA under contract DE-NA0001976.

\bibliography{LSTIbib}

\end{document}